\documentclass[pre,aps,amsmath,amssymb,twocolumn,superscriptaddress,10pt]{revtex4-2}
\usepackage{hyperref}
\hypersetup{
breaklinks=true,
colorlinks=true,
citecolor=blue,
linkcolor=magenta,
filecolor=blue,
urlcolor=blue
}
\usepackage{graphicx}
\usepackage{dcolumn}
\usepackage{bm}

\begin{document}

\preprint{APS/123-QED}

\title{Relativistic Roots of $\kappa$-Entropy}
\author{G. Kaniadakis}
\email{giorgio.kaniadakis@polito.it}
\affiliation{Department of Applied Science and Technology, Politecnico di Torino, \\ Corso Duca degli Abruzzi 24, 10129
Torino, Italy}

\date{\today}

\begin{abstract}
The axiomatic structure of the $\kappa$-statistcal theory is proven. In addition to the first three standard Khinchin--Shannon axioms of continuity, maximality, and expansibility, two further axioms are identified, namely the self-duality axiom and the scaling axiom. It is shown that both the $\kappa$-entropy and its special limiting case, the classical Boltzmann--Gibbs--Shannon entropy, follow unambiguously from the above new set of five axioms. It has been emphasized that the statistical theory that can be built from $\kappa$-entropy has a validity that goes beyond physics and can be used to treat physical, natural, or artificial complex systems. The physical origin of the self-duality and scaling axioms has been investigated and traced back to the first principles of relativistic physics, i.e., the Galileo relativity principle and the Einstein principle of the constancy of the speed of light. It has been shown that the $\kappa$-formalism, which emerges from the $\kappa$-entropy, can treat both simple (few-body) and complex (statistical) systems in a unified way. Relativistic statistical mechanics based on $\kappa$-entropy is shown that preserves the main  features of classical statistical mechanics (kinetic theory, molecular chaos hypothesis, maximum entropy principle, thermodynamic stability, H-theorem, and Lesche stability). The answers that the $\kappa$-statistical theory gives to the more-than-a-century-old open problems of relativistic physics, such as how thermodynamic quantities like temperature and entropy vary with the speed of the reference frame, have been emphasized.
\end{abstract}
\maketitle

\section{Introduction}

The spread of the neologism $\kappa$-distribution within the astrophysical plasma community began after the publication of the seminal paper of Vasyliunas'~\cite{Vasyliunas,KL2004} in 1968. The enormous existing literature on so-called $\kappa$-plasmas shows the undisputed success of the Vasyliunas $\kappa$-distribution, which still seems to be very relevant today. There have been a very high number of attempts to justify it, which shows that none of the proposals put forward are accepted by the whole community of physicists in the field and, therefore, the success of this distribution is mainly of an empirical nature. Curiously, there are no advanced proposals that consider the possibility of going beyond the Vasyliunas $\kappa$-distribution and investigating new possible distributions that have other analytical forms but describe the empirical data equally well.

This paper deals with the statistical theory~\cite{PhA2001, PLA2001} proposed in 2001, which goes beyond the Vasyliunas distribution. The underlying new distribution is also called a $\kappa$-distribution, which sometimes unintentionally causes some confusion for the reader. This choice was made because $\kappa$-plasmas represent one of the most natural fields of applications of the new $\kappa$-distribution. The promotion of the proposal of the new $\kappa$-distribution essentially arises from the purely theoretical need to have a statistical distribution that possesses the important property of self-duality, i.e., $f(-E) f(E) = {\rm constant }$, 
as in the case of the Boltzmann exponential factor $\exp(-\beta E)\exp(\beta E) =1$ of ordinary Boltzmann--Gibbs statistical mechanics. This need was easily met thanks to the empirical evidence suggesting that the Pareto power law trend of the statistical distribution has purely asymptotic validity. The new $\kappa$-distribution asymptotically exhibits a power-law tail but gradually transforms in the intermediate range in its bulk range and takes on the features of the typical behavior of the standard exponential Boltzmann factor. In the two papers published in 2002~\cite{PRE2002} and 2005~\cite{PRE2005}, it was shown that the new $\kappa$-distribution arises naturally in the context of Einstein's special relativity and generates a self-consistent $\kappa$-statistic, which turns out to be a relativistic extension of classical Boltzmann--Gibbs statistical mechanics. The entirety of $\kappa$-statistical mechanics can be traced back to $\kappa$-entropy
\begin{equation}
S_{\kappa}=\sum_{i=1}^W \, \frac{f_i^{\, 1-{\kappa}} - f_i^{\, 1+{\kappa}}}{2 {\kappa}} \ \  \label{20Y-I-1}
\end{equation}
where $\{ f_i \}$ is the statistical distribution. $S_{\kappa}$ entropy is the relativistic generalization of classical Boltzmann--Gibbs--Shannon entropy, which recovers in the $\kappa \rightarrow 0$ limit. The corresponding $\kappa$-distribution behaves like the ordinary Boltzmann distribution at low energies, while it presents a power-law tail at high energies.  
 
One of the greatest successes of $\kappa$-statistics is undoubtedly the explanation of the non-Boltzmannian spectrum of cosmic rays, which are relativistic particles. The persistent power-law tails of this spectrum, spanning 13 decades in terms of energy and 33 decades in terms of particle flux, turn out to be a purely relativistic effect correctly predicted by $\kappa$-statistics.

Remarkably, although the statistical theory based on $S_{\kappa}$ can be traced back to the first principles of special relativity, it can also be introduced without reference to special relativity, as will be shown in Section II, since it also has applications outside relativistic physics. For this reason, statistical theory~\cite{PhA2006,EPJA2009,EPJB2009,EPL2010,PLA2011,MPLB2012,Entropy2013}
based on the $\kappa$-distribution has attracted the interest of many researchers. In the last two decades, various authors have devoted themselves to the study of both the theoretical foundations of the theory and its applications not only in plasma physics but also in various other areas of the science of complex physical, natural, or artificial statistical systems. Some of these works deal with the H-theorem and the molecular chaos hypothesis
\cite{Silva06A,Silva06B}, thermodynamic stability
\mbox{\cite{Wada1,Wada2}}, Lesche stability
\mbox{\cite{KSPA04,AKSJPA04,Naudts1,Naudts2}}, the Legendre structure of the resulting thermodynamics~\cite{ScarfoneWada,Yamano},
the thermodynamics of non-equilibrium systems~\cite{Lucia2010},
quantum versions of the theory~\cite{AlianoKM2003,Santos2011a,Santos2011b,Santos2012},
the geometric structure of the theory~\cite{Pistone,Pistone_Shoaib_2023},
various mathematical aspects of the theory
\cite{KLS2004,KLS2005,KSphysA2003,Oikonomou2010,Stankovic2011,Tempesta2011,DeossaCasas,Vigelis,Scarfone2013,Biro_2022_entropy,Sfetcu_et_al_2022_Entropy,Sfetcu_et_al_2022_MPAG,Wada_Scarfone_2023_entropy,Scarfone_Wada_2024_entropy}, etc. On the other hand, specific applications to physical systems have been
considered, e.g., cosmic rays~\cite{PRE2002}, relativistic
\cite{GuoRelativistic} and classical~\cite{GuoClassic} plasmas  in presence of external electromagnetic fields, relaxation in relativistic plasmas under wave--particle interactions
\cite{Lapenta,Lapenta2009}, electronic cooling~\cite{daSilva2021PhysicaANewton}, dark energy models~\cite{Rani_et_al_2022_IJMPD31,Ghaffari_2022_MPLA,Sharma_et_al_2022_IJMPD31,Drepanou_et_al_2022_EPJC,Korunur_2022_IJMPA,HernandezAlmada_et_al_2022a,HernandezAlmada_et_al_2022b,Blasone_et_al_2023,Sania_et_al_2023,Dubey_et_al_2023,Jawad_et_al_2023,Singh_et_al_2023,Kumar_et_al_2023,Sharma_et_al_2023,Sadeghi_et_al_2023,Kumar_et_al_2024_NewAstronomy,Sultana_Chattopadhyay_2024,Chokyi_Chattopadhyay_2024_AnnPhys,Ali2024,Rao2024,YarahmadiEPJC2024}, quantum gravity~\cite{Abreu_et_al_2016_APL114,Abreu_et_al_2017,Chen_et_al_2017_CPL34,Abreu_et_al_2018,Abreu_et_al_2018EPL124,Abreu_et_al_2019,Yang_et_al_2022,He_2022}, quantum cosmology~\cite{Moradpour_et_al_2022_Entropy,Luciano_2022_EPJC,Luciano_Saridakis_2023Pv,Lambiase_et_al_2023,Sheykhi2024}, gravitation and cosmology~\cite{Luciano2022Entropy,Sadeghnezhad2023}, anomalous diffusion
\cite{WadaScarfone2009,Wada2010}, non-linear kinetics~\cite{KQSphysA2003,BiroK2006,Casas,Hirica_et_al_2022_Mathematics,Gomez_et_al_2023_FPE,Evangelista_Lenzi_2023_EntropySI,Guha_2023_EntropySI},
the kinetics of interacting atoms and
photons~\cite{Rossani}, particle kinetics in the presence of
temperature gradients~\cite{GuoDuoTgradient,Guo2012},  particle systems in
external conservative force fields~\cite{Silva2008}, stellar
distributions in astrophysics~\cite{Carvalho,Carvalho2,Carvalho2010,Bento2013},
quark--gluon plasma formation~\cite{Tewel}, quantum hadrodynamics models
\cite{Pereira}, fracture propagation~\cite{Fracture}, plasma physics~\cite
{Gougam_Tribeche_2016,Lourek_Tribeche_2016,Lopez_et_al_2017,Chen_et_al_2017,Saha_Tamang_2017,Lourek_Tribeche_2019,Khalid_Rahman_2020,Tan_et_al_2022,Irshad_et_al_2022_EPJPLUS,Bellahsene_et_al_2023,Dubinov_2023,Raut_et_al_2023,Irshad_et_al_2023,Khalid_et_al_2023,Bala_Kaur_2024}, seismology~\cite{Hristopulos_et_al_2014_PRE89,Hristopulos_et_al_2015_Entropy17,daSilva2021ChaosSolFractal,Hristopulos_Baxevani_2022_Entropy24}, seismic imaging~\cite{daSilva2020PRE,daSilva2021bPhysicaA,daSilva2022PRE,daSilva2023Entropy,daSilva_Kaniadakis2023SEG}, nuclear physics~\mbox{\cite{deAbreu_et_al_2019,deAbreu_Martinez_2020,deAbreu_et_al_2022_Entropy,deAbreu_et_al_2022_NET,Martinez_deAbreu_2023}}, and quantum mechanics~\cite{Portesi,Korea,PRD}. Other
applications concern dynamical systems at the edge of chaos~
\cite{Corradu,Tonelli,Celikoglu}, fractal systems~\cite{Olemskoi},
field theories~\cite{Olemskoi2010}, genomic analysis~\cite{Souza_et_al_2014_EPL,Costa_et_al_2019_PRE,deLima_et_al_2022_EntropySI},
random matrix theory~\cite{AbulMagd,AbulMagd2009,AbulMagd2012}, robust statistical inference~\cite{daSilva2021EPJplus,dosSantosLima2023PlosOne},
 error theory~\cite{WadaSuyari06,daSilva2022PhysicaA}, game theory~\cite{Topsoe}, the theory of complex networks~\cite{Macedo},
information theory~\cite{WadaSuyari07}, etc. Also, applications to
economic systems have been considered, e.g., to study the personal
income distribution~\cite{Clementi,Clementi2008,Clementi2009,Clementi2011,Clementi2012a,Clementi2012b,Clementi2023}, to model deterministic heterogeneity in tastes and product
differentiation~\cite{Rajaon,Rajaon2008}, in finance~\cite{Trivellato2012,Trivellato2013}, in equity options~\cite{Tapiero}, to construct  taxation and redistribution models~\cite {Bertotti}, etc.

In this paper, we present some new aspects of $\kappa$-statistical theory. Section \ref{sec2} focuses on the axiomatic structure of the theory by proposing the five axioms from which the theory can be deduced without referring to the principles of special relativity. Section \ref{sec3} focuses on the relativistic origin of the theory. Some peculiar aspects of the physical--mathematical formalism of the theory are emphasized and, in particular, it is shown how the axioms of the theory emerge in relativistic physics. Finally, in Section \ref{sec4}, a synthetic overview of the theory is given in the light of the results obtained in recent years.

\section{An Axiomatic Approach to \boldmath{$\kappa$}-Entropy}\label{sec2}

The concept of entropy was introduced in the second half of the nineteenth century in the context of thermodynamics by Clausius, who also gave it its name, and immediately afterward by Boltzmann in the context of statistical mechanics. This physical quantity, which emerged within the framework of classical physics, has retained its original form over time, even after the emergence of new branches of physics such as relativistic physics and quantum physics. This entropy, which is still used in physics, was also introduced towards the middle of the twentieth century in Shannon's information theory and subsequently in various fields of science to treat physical, natural, or artificial complex systems. Currently, this entropy is called Boltzmann--Gibbs--Shannon (BGS) entropy~\cite{Shannon,Khinchin} and is a special case of the more general class of the trace form generalized entropic functional 
\begin{equation}
S=\sum_{i=1}^W \sigma(f_i)=-\sum_{i=1}^W f_i \Lambda (f_i)= -<\Lambda (f_i)> \ \   \label{20Y-II-2}
\end{equation}
where $<>$ indicates the standard mean value, and in the distribution $f=\{f_1, f_2, ..., f_i, ..., f_W\}$, $f_i$ represents the probability that the system is in the microstate $i$ with $\sum_1^W f_i =1$. The standard BGS entropy is obtained by setting $\Lambda(f_i)=\ln(f_i)$. In expression (\ref{20Y-II-2}) of the generalized entropy~\cite{Csiszar,Zografos}, the function $\Lambda(f_i)$, called the generalized logarithm, is an arbitrary strictly increasing function that is negative on the interval $0<f_i<1$. The function $\sigma (f_i)=-f_i \, \Lambda (f_i)$ represents the contribution to entropy associated with the state $i$.

Some meaningful properties of the BGS entropy that are elevated to the rank of axioms~\cite{Shannon,Khinchin}, i.e., the Khinchin--Shannon (KS) axioms I, II, and III, can also apply to the generalized entropies. It is therefore assumed that the generalized entropy defined in Equation~$(\ref{20Y-II-2})$ obeys the following three KS axioms:

 \noindent
{ {I. Continuity axiom:}
} The entropy depends continuously on all the variables $f_i$. From this axiom follows the continuity of the function $\Lambda(p_i)$.

 \noindent
{{II. Maximality axiom:}} The entropy is maximized by the uniform distribution $f_W=\{f_1=\frac{1}{W}, f_2=\frac{1}{W}, ..., f_i=\frac{1}{W}, ..., f_W=\frac{1}{W}\}$, i.e., $S[f]\leq S[f_W]$. From this axiom follows the concavity property $\frac{d^2 \, \sigma(f_i)}{df_i^2}<0$.

 \noindent
{ {III. Expansibility axiom:}} The $(W+1)$-component distribution $g$ obtained after the expansion of the $W$-component distribution $f$ by adding a component with probability equal to zero corresponds to the same entropy of the distribution $g$, i.e., $S[g]=S[f]$. From this axiom follows the property  $0^+\Lambda(0^+)=0$. We also recall that the particular probability distribution $f=\{ \delta_{ia}, \, 1\leq i \leq W \}$, where $a$ is a given integer with $1\leq a \leq W$, describes a state for which one has the maximum information. For this state, $S=0$ must be set. This condition in turn states that $0^+\Lambda(0^+)=0$ and also that $\Lambda(1)=0$. Equivalently, we can set up $\sigma(0)=\sigma(1)=1$.

It is noteworthy that although the above three KS axioms impose some properties on the function $\Lambda(f_i)$ and then on $\sigma(f_i)$, they do not uniquely determine its form. In the case of BGS entropy, the form of the function $\Lambda(f_i)$ is determined by the fourth KS axiom, i.e., the separability or strong additivity axiom, which implies the property $\Lambda(f_i\, g_j)=\Lambda(f_i)+\Lambda(g_j)$, from which $\Lambda(f_i)=\ln(f_i)$ follows. To go beyond the logarithmic BGS entropy and introduce new entropic functionals, it is necessary to abandon the fourth KS axiom, provided that the first three KS axioms are still equally valid. The fourth KS axiom is replaced by two meaningful properties of the BGS entropy that can equally define the BGS entropy form without invoking the additivity property of the ordinary logarithm function. These two properties are elevated to the status of new axioms and must also apply to the case of generalized entropies. The problem is, therefore, reduced to the search for new generalized entropies that, in addition to the ordinary BGS entropy, also obey the two new~axioms. 

 Starting from the generalized logarithm $\Lambda(f_i)$, we introduce the function $\Lambda(1/f_i)$, which we will call generalized surprise or generalized unexpectedness in analogy to the terms surprise~\cite{Watanabe} or unexpectedness~\cite{Barlow} used in the literature when the generalized logarithm is reduced to the ordinary logarithm. The generalized surprise/unexpectedness is a continuous, decreasing function that admits a unique zero at $f_i=1$. The opposite of the generalized surprise/unexpectedness $\Lambda^*(f_i)=-\Lambda(1/f_i)$ is a continuous, increasing function and is referred to below as the dual generalized logarithm. The two generalized logarithms $\Lambda(f_i)$ and $\Lambda^*(f_i)$, are the duals of each other and are both increasing functions on the interval $0<f_i<+\infty$ with a zero at $f_i=1$. The two functions $\sigma(f_i)=-f_i\, \Lambda(f_i)$ and $\sigma^*(f_i)=-f_i\, \Lambda^*(f_i)$ can be employed to construct the two entropic functionals $S=\sum_i \sigma(f_i)$ and $S^*=\sum_i \sigma^*(f_i)$, respectively, both of which fulfill the first three KS axioms. In general, $S^*\neq S$ holds, and this leads to a theoretical dichotomy, since the two entropies $S$ and $S^*$ define two different statistical theories and, most worryingly, there is no criterion for choosing one of the two entropies. This dilemma does not exist in the case of ordinary BGS entropy, because the property $\ln(1/f_i)=-\ln(f_i)$ implies the self-duality of the logarithm $\ln^*(f_i)=\ln(f_i)$ and then the self-duality of the entropy, i.e., $S^*=S$. To guarantee the uniqueness of the entropy form when considering a generalized statistical theory, we must force the generalized logarithm to be self-dual, just as in the case of ordinary statistical theory. Then,we can introduce the following axiom:

\noindent
{ {IV. Self-duality axiom:}} The entropy defined in Equation~(\ref{20Y-II-2}) must be considered both as the standard mean value of the opposite of the generalized logarithm $-\Lambda(f_i)$ and as the standard mean value of the generalized surprise/unexpectedness $\Lambda(1/f_i)$, i.e.,
\begin{equation}
S=- <\Lambda \left (f_i \right ) > \,\, = \, \, <\Lambda \left ( 1/f_i \right ) > \ \  \label{20Y-II-3}
\end{equation}
or, equivalently, the generalized logarithm must possess the self-duality property
\begin{equation}
\Lambda (1/f_i)=-\Lambda (f_i) \ \  \label{20Y-II-4}
\end{equation}

It is noteworthy that axioms I, III, and IV concern some properties of the function $\sigma (f_i)$ or equivalently of $\Lambda(f_i)$, while axiom II concerns a precise property of the function $\frac{ d^2 \sigma(f_i)}{d f_i^2}$. In the following, we will focus on a property of the function $\lambda(f_i)=-\frac{d \sigma(f_i)}{d f_i}= \frac{d }{d f_i} f_i \Lambda (f_i) $. First, recall that $\sigma(f_i)$ is a continuous and concave function with $\frac{ d^2 \sigma(f_i)}{d f_i^2}<0$, which has two zeros $\sigma(0)=\sigma(1)=0$. Then, $\sigma(f_i)$ presents its maximum value at $f_i=1/\epsilon$ with $\epsilon > 1$. This means that $\lambda(f_i)$ is a monotonically increasing function that has a zero at $f_i=1/\epsilon$. These general features of the function $\lambda(f_i)$ are typical of a generalized logarithm, with the exception that the generalized logarithm has its zero at $f_i=1$. Recall that in the case of the ordinary logarithm, the associated function $\lambda(f_i)$ is simply the ordinary logarithm after it has been properly scaled, i.e., $\lambda(f_i)=\frac{1}{\gamma} \ln(\epsilon f_i)$, with $\gamma=1$ and $\epsilon=e$ ($e$ is the Napier number). This scaling property of the ordinary logarithm must also apply to the generalized logarithm $\Lambda(f_i)$, so that the relationship $\lambda(f_i)=\frac{1}{\gamma} \Lambda(\epsilon f_i)$ must hold, with $\gamma$ and $\epsilon$ being two scaling parameters that are connected by the Boltzmann limit $\lim_{\gamma \rightarrow 1}\epsilon=e$. The above scaling property of $\Lambda(f_i)$ is elevated to the status of the following axiom:

\noindent 
{{V. Scaling axiom:}} The generalized logarithm which appears in the definition of entropy (\ref{20Y-II-2}) has the following property of scaling:
\begin{equation}
\frac{d}{d f_i}\big(f_i \Lambda (f_i)\big)= \frac{1}{\gamma} \Lambda (\epsilon f_i) \ \  \label{20Y-II-5}
\end{equation}
where $\gamma$ and $\epsilon$ are the scaling parameters.

The question naturally arises as to whether the BGS entropy is the only existing entropy that obeys the two axioms of self-duality and scaling or whether there is another generalized entropy that equally fulfills the two axioms mentioned. To answer this question, we start from Equation~(\ref{20Y-II-5}), which expresses the scaling axiom and is to be regarded as a differential--functional equation. We seek its general solution after we have correctly determined the free scaling parameters $\gamma$ and $\epsilon$.  Equation~(\ref{20Y-II-5}) was solved in~\cite{PRE2002,KLS2005}, and it was shown that besides the BGS entropy, there is a large class of entropies obeying the scaling axiom, some of which are already known in the literature~\cite{PRE2002,KLS2005}. This class of generalized entropies is drastically reduced if the generalized entropy must simultaneously satisfy the scaling and self-duality axioms. In this case, the above class of generalized entropies is reduced to only two entropies. Only the standard BGS entropy corresponding to $\Lambda(f_i)=\ln(f_i)$ and the so-called $\kappa$-entropy $S_{\kappa}$ corresponding to the $\kappa$-logarithm $\Lambda(f_i)=\ln_{\kappa}(f_i)$ remain to obey the two scaling and self-duality axioms simultaneously.  The $\kappa$-logarithm is defined by
\begin{equation}
\ln_{\kappa}(f_i) = \frac{f_i^{\,\kappa}-f_i^{-\kappa}}{2 \kappa}= \frac{1}{\kappa} \sinh (\kappa \ln (f_i))  \ \   \label{20Y-II-6}
\end{equation}
{The} free parameter that appears in the expression of the $\kappa$-logarithm varies in the range of $0<\kappa<1$ and in the $\kappa \rightarrow 0$ limit,  the $\kappa$-logarithm $\ln_{\kappa}(f_i)$ is reduced to the ordinary logarithm $\ln(f_i)$. The function $\ln_{\kappa}(f_i)$ can then be regarded as a one-parameter generalization of the ordinary logarithm. Remarkably, the meaning of the parameter $\kappa$ emerges when the asymptotic behaviour of the $\kappa$-logarithm is considered. The asymptotic behaviour of the $\kappa$-logarithm results from Equation~(\ref{20Y-II-6}), i.e., for $f_i \rightarrow 0^+$, it obtains $\ln_{\kappa}(f_i) \propto -f_i^{-\kappa}$, while for $f_i \rightarrow +\infty$, according to self-duality axiom, it results in $\ln_{\kappa}(f_i) \propto f_i^{\,\kappa}$. The parameter $\kappa$ turns out to be the Pareto index, which characterizes the power-law asymptotic behavior of the $\kappa$-logarithm. Finally, the constants $\gamma$ and $\epsilon=\exp_{\kappa}(\gamma)$ are given by
\begin{equation}
\gamma = \frac{1}{\sqrt{1-\kappa^2}} \ \  \ , \ \ \ \epsilon = \left (\frac{1+\kappa}{1-\kappa} \right )^{\frac{1}{2\kappa}} \ \   \label{20Y-II-7}
\end{equation}
and in the $\kappa \rightarrow 0$ limit, they reduce to unity and Napier number $e$, respectively, reproducing the results of the standard logarithmic entropy. The connection between the parameters $\gamma$ and  
 $\epsilon$ follows directly from their expressions and is given by  $\gamma=\ln_{\kappa}(\epsilon)$. 

Besides the BGS entropy, the $\kappa$-entropy is the only one that simultaneously fulfills all five axioms presented above. Thanks to the self-duality property of the $\kappa$-logarithm, i.e., $\ln_{\kappa}(1/f_i)=-\ln_{\kappa}(f_i)$, $\kappa$-entropy can be written as follows:
\begin{equation}
S_{\kappa}= \sum_{i}^W \sigma_{\kappa}(f_i) =- \sum_{i}^W f_i \ln_{\kappa}(f_i) = \sum_{i}^W f_i \ln_{\kappa}(1/ f_i)  \ \   \label{20Y-II-8}
\end{equation}
and can be regarded as the standard mean of both the opposite of the $\kappa$-logarithm and its self-dual $\kappa$-surprise/unexpectedness i.e.  
$S_{\kappa}=- <\ln_{\kappa} \left (f_i \right ) > \,\, = \, \,  <\ln_{\kappa} \left ( 1/f_i \right ) >$.

This axiomatic approach to the introduction of $\kappa$-entropy is typical of information theory. Remarkably, the two self-duality and scaling axioms that give rise to $\kappa$-entropy are also valid in the framework of BGS entropy, although they do not have the rank of axioms but rather express two important properties of standard entropy. The method of replacing the strong additivity axiom of BGS entropy with the new self-duality and scaling axioms that do not contradict any of the standard properties of BGS entropy, including its additivity, clearly leads to a new generalized entropy, namely $\kappa$-entropy, which can be employed to analyze physical or non-physical complex systems.

In the reference~\cite{IKGS2021}, about sixty different entropies are given, and the corresponding list is not complete. For each of these generalized entropies, it is in principle possible to identify the founding axioms that follow the standard lines of information theory~\cite{Csiszar}. In any case, it is important to emphasize that entropy is a physical concept that was first introduced in the context of classical thermodynamics and statistical physics. This means that a generalized entropy that claims to be physically meaningful should not only be introduced by postulating some mathematical axioms, as we have done here, but also that these founding axioms should emerge within the framework of a physical theory, starting from its first principles.

The task of the next section will be to show that the two axioms of self-duality and scaling, as well as the $\kappa$-logarithm form, follow naturally from the first principles of special~relativity.

\section{Special Relativity}\label{sec3}

\subsection{Energy--Momentum Lorentz Transformations}

Let us consider two identical particles $A$ and $B$ with rest mass $m$ in the one-dimensional inertial frame $\cal S$, whose velocities, momenta, and total energies are given by $v_{\scriptscriptstyle A}$, $p_{\scriptscriptstyle
A}=mv_{\scriptscriptstyle A}\gamma(v_{\scriptscriptstyle A})$, and $E_{\scriptscriptstyle
A}=mc^2\gamma(v_{\scriptscriptstyle A})$ and $v_{\scriptscriptstyle A}$, $p_{\scriptscriptstyle
B}=mv_{\scriptscriptstyle B}\gamma(v_{\scriptscriptstyle B})$, and $E_{\scriptscriptstyle
B}=mc^2\gamma(v_{\scriptscriptstyle B})$, respectively, where $\gamma(v_{\scriptscriptstyle})=(1-v_{\scriptscriptstyle
}^2/c^2)^{-1/2}$ is the Lorentz factor, and $c$ is the light speed. 

In the rest frame $\cal S'$ of particle $B$, the above variables in the case of particle $B$ assume the values  $v'_{\scriptscriptstyle B}=0$,
$p'_{\scriptscriptstyle B}=0$, and $E'_{\scriptscriptstyle B}=mc^2$, respectively, while the velocity $v'_{\scriptscriptstyle A}$ of particle $A$ is given by the Einstein relativistic velocity
additivity law $v'_{\scriptscriptstyle A}=(v_{\scriptscriptstyle
A}-v_{\scriptscriptstyle B})/(1-v_{\scriptscriptstyle A}
v_{\scriptscriptstyle B}/c^2)$. In the same frame $\cal S'$,  the momentum $p'_{\scriptscriptstyle A}$ and the energy $E'_{\scriptscriptstyle A}$ of the particle $A$ are given by the dynamic Lorentz transformations
\begin{eqnarray}
&&p'_{\scriptscriptstyle A}=\gamma( v_{\scriptscriptstyle B})
p_{\scriptscriptstyle A}-c^{-2}v_{\scriptscriptstyle B}\gamma(
v_{\scriptscriptstyle B})E_{\scriptscriptstyle A} \ 
\label{20Y-III-9} \\
&&E'_{\scriptscriptstyle A}=\gamma( v_{\scriptscriptstyle B})
E_{\scriptscriptstyle A}-v_{\scriptscriptstyle B}\gamma(
v_{\scriptscriptstyle B})p_{\scriptscriptstyle A} \  \label{20Y-III-10}
\end{eqnarray}
{After} introducing the momentum $p_{\scriptscriptstyle
B}=mv_{\scriptscriptstyle B}\gamma(v_{\scriptscriptstyle B})$ and the energy $E_{\scriptscriptstyle
B}=mc^2\gamma(v_{\scriptscriptstyle B})$ of the particle $B$, the above transformations assume the form
\begin{eqnarray}
&&p'_{\scriptscriptstyle A}=\frac{1}{mc^2} p_{\scriptscriptstyle A}
E_{\scriptscriptstyle B}-\frac{1}{mc^2} E_{\scriptscriptstyle A}
p_{\scriptscriptstyle B} \ 
\label{20Y-III-11} \\
&&E'_{\scriptscriptstyle A}= \frac{1}{mc^2} E_{\scriptscriptstyle A}
E_{\scriptscriptstyle B}-\frac{1}{m} p_{\scriptscriptstyle A}p_{\scriptscriptstyle B} \  \label{20Y-III-12}
\end{eqnarray}

It will be more useful for our discussion hereafter to introduce the new dimensionless variables
$(u,q,{\cal E})$ in place of the dimensional variables $(v,p,E)$ through 
\begin{eqnarray}
\frac{v}{u}=\frac{p}{mq}=\sqrt{\frac{E}{m{\cal E}}}=\kappa c =v_*<c \ \  \label{20Y-III-13}
\end{eqnarray}
where $v_*$ is an arbitrary reference velocity.  For a particle at rest, this results in $E(0)=m\,c^2$ and then ${\cal E}(0)=1/\kappa^2$ so that $1/\kappa^2$ is the dimensionless rest energy of the particle. Finally, we note that the classical $c\rightarrow \infty$ limit is replaced now by the
 $\kappa\rightarrow 0$ limit.  
 
The Lorentz transformations for the dimensionless momentum and energy variable $q$ and ${\cal E}$ assume the form
\begin{eqnarray}
&&q'_{\scriptscriptstyle A}=\kappa^2 q_{\scriptscriptstyle A} {\cal
E}_{\scriptscriptstyle B} - \kappa^2 q_{\scriptscriptstyle B} {\cal
E}_{\scriptscriptstyle A} \ 
\label{20Y-III-14} \\
&&{\cal E}'_{\scriptscriptstyle A}=\kappa^2 {\cal
E}_{\scriptscriptstyle A} {\cal E}_{\scriptscriptstyle B} -
q_{\scriptscriptstyle A} q_{\scriptscriptstyle B} \  \label{20Y-III-15}
\end{eqnarray}

\subsection{Emergence of $\kappa$-Exponential Function in Special Relativity}

By directly combining the Lorentz transformations, we obtain
\begin{eqnarray}
\kappa^2{\cal E}'_{\scriptscriptstyle A} \pm \kappa q'_{\scriptscriptstyle A}=\left (\kappa^2{\cal E}_{\scriptscriptstyle B} \mp \kappa q_{\scriptscriptstyle B} \right ) \! \left ( \kappa^2{\cal E}_{\scriptscriptstyle A} \pm \kappa q_{\scriptscriptstyle A} \right ) \ \ 
\label{20Y-III-16} 
\end{eqnarray}
The variables $\kappa^2{\cal E} \pm \kappa q \geq 0$  can be viewed as a dynamic light cone variable, and Equation~(\ref{20Y-III-16}) can be written in the form
\begin{equation}
\left ( \kappa^2{\cal E}'_{\scriptscriptstyle A} \pm \kappa q'_{\scriptscriptstyle A} \right)^{1/\kappa}=\left (\kappa^2 {\cal E}_{\scriptscriptstyle B} \mp \kappa q_{\scriptscriptstyle B} \right )^{1/\kappa} \! \left ( \kappa^2{\cal E}_{\scriptscriptstyle A} \pm \kappa q_{\scriptscriptstyle A} \right )^{1/\kappa} \ \ 
\label{20Y-III-17} 
\end{equation}
After taking into account that $\lim_{\kappa \rightarrow 0} \left (\kappa^2{\cal E} \pm \kappa q \right )^{1/\kappa}=\exp(\pm q)$, the latter relationship reduces to  $\exp(\pm q'_{\scriptscriptstyle A})=\exp(\mp q_{\scriptscriptstyle B}) \exp(\pm q_{\scriptscriptstyle A})$, which implies the Galilei transformations for the momenta $q'_{\scriptscriptstyle A} = q_{\scriptscriptstyle A}
-\,q_{\scriptscriptstyle B}$. This result suggests the introduction of the new variable
\begin{eqnarray}
\exp_{\kappa}(q)= \left(\kappa^2 {\cal E} + \kappa\,q\right )^{1/\kappa} \  \label{20Y-III-18}    
\end{eqnarray}
generalizing within the special relativity the ordinary exponential $\exp(q)$, which recovers in the classical limit, i.e., $\lim_{\,\kappa \rightarrow 0}\, \exp_{\kappa}(q)= \exp(q)$. The Lorentz transformations, as given by Equation~(\ref{20Y-III-17}), in terms of the $\kappa$-exponential function,  assume the form
\begin{eqnarray}
\exp_{\kappa}(\pm q'_{A})=\exp_{\kappa}(\mp q_{ B})\,\exp_{\kappa}(\pm q_{A}) \ \ 
\label{20Y-III-19}
\end{eqnarray}

Starting from the Lorentz transformations (\ref{20Y-III-14}) and (\ref{20Y-III-15}), the relationship expressing the Lorentz invariance, i.e., $\kappa^4{\cal E}'^2-\kappa^2q'^2=\kappa^4{\cal E}^2-\kappa^2q^2$, can be obtained, and after identifying ${\cal S}'$ as the particle rest frame where ${\cal E}(0)=1/\kappa^2$, the energy--momentum dispersion relation 
\begin{eqnarray}
\kappa^4{\cal E}^2-\kappa^2q^2=1 \  \label{20Y-III-20}
\end{eqnarray}
follows. From the latter relationship, the expression of the dimensionless total energy ${\cal E}$ can be obtained in terms of the dimensionless momentum $q$ 
\begin{eqnarray}
{\cal E} =\frac{1}{\kappa^2}\sqrt{ 1+\kappa^2\,q^2}  \ 
\label{20Y-III-21} 
\end{eqnarray}
{After} inserting this expression of total energy in the definition (\ref{20Y-III-18}) of the $\kappa$-exponential function, the explicit form is obtained as follows:
\begin{equation}
\exp_{\kappa}(q) = \left ( \sqrt{1+ \kappa^2 q^2} + \kappa q \right )^{1/\kappa}= \exp \left ( \frac{1}{\kappa} {\rm arcsinh} (\kappa q) \right )  \ \   \label{20Y-III-22}
\end{equation}

After the substitution of the expression of the dimensionless total energy given by Equation~(\ref{20Y-III-21}) into the first of the Lorentz transformations given by Equation~(\ref{20Y-III-14}), the relativistic additivity law of the dimensionless momenta assumes the form $q'_{\scriptscriptstyle A}=q_{\scriptscriptstyle A}\stackrel{\kappa}{\ominus}q_{\scriptscriptstyle B}=q_{\scriptscriptstyle A}\stackrel{\kappa}{\oplus}(-q_{\scriptscriptstyle B})$, where the $\kappa$-sum $\stackrel{\kappa}{\oplus}$ is defined as
\begin{eqnarray}
q_{\scriptscriptstyle A}\stackrel{\kappa}{\oplus}q_{\scriptscriptstyle B}= q_{\scriptscriptstyle A} \sqrt{ 1+\kappa^2\,q_{\scriptscriptstyle B}^2} + q_{\scriptscriptstyle B} \sqrt{ 1+\kappa^2\,q_{\scriptscriptstyle A}^2} \ 
\label{20Y-III-23} 
\end{eqnarray}

The following property of the $\kappa$-exponential holds:
\begin{eqnarray}
\exp_{\kappa}(q_{\scriptscriptstyle A}\stackrel{\kappa}{\oplus}q_{\scriptscriptstyle B}) = \exp_{\kappa}( q_{A}) \exp_{\kappa}(q_{B}) \ \ 
\label{20Y-III-24}
\end{eqnarray}
which is reminiscent of the analogous property of the classical exponential function $\exp (q_{\scriptscriptstyle A}+q_{\scriptscriptstyle A})= \exp (q_{\scriptscriptstyle A}) \, \exp (q_{\scriptscriptstyle B})$.

\subsection{Emergence of $\kappa$-Logarithm Function in Special Relativity}

The function $\ln_{\kappa}(w)$ is defined as the inverse function
of $\exp_{\kappa}(w)$ through 
$\ln_{\kappa}(\exp_{\kappa}w)=\exp_{\kappa}(\ln_{\kappa}w)=w$.  Its explicit expression is 
\begin{eqnarray}
\ln_{\kappa}(w) = \frac{w^{\kappa}-w^{-\kappa}}{2\kappa} = \frac{1}{\kappa}\,\sinh \, (\kappa \ln w)\ \ 
\label{20Y-III-25}
\end{eqnarray}
and it reduces to the ordinary logarithm in the classical limit, i.e., $\lim_{\kappa \rightarrow 0}\ln_{\kappa}(w)=\ln
(w)$.

The ordinary logarithm $\ln(w)$ is the only existing function unless a
multiplicative constant is used, which results in the solution to the function equation $\ln(w_1w_2)=\ln(w_1)+\ln(w_2)$. 
Let us now consider the relativistic generalization of this equation, which we obtain from the Lorentz transformation given by Equation~(\ref{20Y-III-24}) after posing $w=\exp_{\kappa}(q)$, i.e.,  $\ln_{\kappa}(w_1 w_2)= \ln_{\kappa}(w_1)\stackrel{\kappa}{\oplus} \ln_{\kappa}(w_2)$, which is written as
\begin{equation}
\ln_{\kappa}(w_1 w_2)= \ln_{\kappa}(w_1)\,\gamma_{\kappa}(\ln_{\kappa}(w_2))
+ \ln_{\kappa}(w_2)\,\gamma_{\kappa}(\ln_{\kappa}(w_1)) \ \ \  \label{20Y-III-26}
\end{equation}
where $\gamma_{\kappa}(\ln_{\kappa}(w))=\sqrt{1+\kappa^2\,\ln_{\kappa}^2(w)}$  is the Lorentz factor of argument $\ln_{\kappa}(w)$. 
By the direct substitution of the $\kappa$-logarithm in the expression of $\gamma_{\kappa}(\ln_{\kappa}(w))$, the further expression $\gamma_{\kappa}(\ln_{\kappa}(w))=(w^{\kappa}+w^{-\kappa})/2$ is obtained. Starting from this latter relationship and after some tedious but straightforward calculation, a third expression is obtained of the function $\gamma_{\kappa}(\ln_{\kappa}(w))$, i.e., 
\begin{eqnarray}
\gamma_{\kappa}(\ln_{\kappa}(w))
= \frac{1}{\gamma} \,\ln_{\kappa}\left(\epsilon w\right) - \ln_{\kappa}\left( w\right)\ \ \  \label{20Y-III-27}
\end{eqnarray}
where the constant $\epsilon=((1+\kappa)/(1-\kappa))^{1/2\kappa}$ represents the $\kappa$-generalization of the Napier number $e$, while the constant $\gamma=1/\sqrt{1-\kappa^2}$ is the Lorentz factor corresponding to the reference velocity $v=v_*$. The two constants are linked through $\epsilon=\exp_{\kappa}(\gamma)$. Equation~(\ref{20Y-III-27}) expresses an important property of the $\kappa$-logarithm, which will be used in the following.

It is noteworthy that the introduction of the function $\ln_{\kappa}(w)$ allows us to write the additivity law of dimensionless relativistic moments defined in Equation~(\ref{20Y-III-24}) in the form
\begin{eqnarray}
q_{\scriptscriptstyle A}\stackrel{\kappa}{\oplus}q_{\scriptscriptstyle B} = \ln_{\kappa}\!\big(\exp_{\kappa}( q_{A}) \exp_{\kappa}(q_{B})\big) \ \ 
\label{20Y-III-28}
\end{eqnarray}

\subsection{Emergence of Self-Duality in Special Relativity}

The dispersion relation (\ref{20Y-III-20}) can be written in the factorized form 
$\left(\kappa^2 {\cal E} + \kappa\,q\right )\left(\kappa^2 {\cal E} - \kappa\,q\right )=1$ and after noticing that $\kappa^2 {\cal E} \pm \kappa\,q \geq 0$, the dispersion relation can be rewritten as follows:
\begin{eqnarray}
\left(\kappa^2 {\cal E} + \kappa\,q\right )^{1/\kappa}\left(\kappa^2 {\cal E} - \kappa\,q\right )^{1/\kappa}=1 \ 
\label{20Y-III-29} 
\end{eqnarray}
and finally, after involving the $\kappa$-exponential function, the relation can be rewritten as 
\begin{eqnarray}
\exp_{\kappa}(q)\,\exp_{\kappa}(-q)=1 \ \ 
\label{20Y-III-30}
\end{eqnarray}
{The} latter relationship expresses an important property of the $\kappa$-exponential function, which, in the classical limit, is reduced to the well-known property of the ordinary exponential function $\exp(q)\,\exp(-q)=1$. As in the case of the ordinary exponential function, the values of the $\kappa$-exponential function for $q<0$ are directly related to its values for $q>0$, resulting in $\exp_{\kappa}(-q)=1/\exp_{\kappa}(q)$. This self-duality property in terms of the $\kappa$-logarithm assumes the form
\begin{eqnarray}
\ln_{\kappa}(1/w)=-\ln_{\kappa}(w) \ \ 
\label{20Y-III-31}
\end{eqnarray}
and means that the values of the $\kappa$-logarithm function on the interval $w>1$ are related to its values on the interval $0<w<1$. 
An important consequence of the relationship (\ref{20Y-III-30}) is that the inverse transformations of the direct Lorentz transformations (\ref{20Y-III-19}) assume the form
\begin{eqnarray}
\exp_{\kappa}(\pm q_{A})=\exp_{\kappa}(\pm q_{ B})\,\exp_{\kappa}(\pm q'_{A}) \ \ 
\label{20Y-III-32}
\end{eqnarray}
{A} comparison of the direct (\ref{20Y-III-19}) and inverse (\ref{20Y-III-32}) Lorentz transformations shows that the inverse Lorentz transformations have the same structure as the direct transformations, except for the substitutions
$q'_{\scriptscriptstyle A} \leftrightarrow q_{\scriptscriptstyle A}$
and $q_{\scriptscriptstyle B} \rightarrow - q_{\scriptscriptstyle
B}$. This symmetry expresses the Galilean principle of relativity, which applies both in classical physics and in special relativity and prescribes the equivalence of all inertial frames. From this, we can conclude that the self-duality property $\exp(q) \exp(-q)=1$ of the ordinary exponential function and the analogous property of the $\kappa$-exponential function, which is given by Equation~(\ref{20Y-III-30}), is enforced by the Galilean principle of relativity.

\subsection{$\kappa$-Mathematics}

The additivity law of dimensionless relativistic moments defined in Equation~(\ref{20Y-III-23}) with $q_A, q_B \in {\bf R}$ called $\kappa$-sum and denoted by $\stackrel{\kappa}{\oplus}$ is a generalized sum and can be viewed as a one-parameter, continuous deformation of the ordinary sum, which recovers in the classical limit
$\kappa\rightarrow 0$, i.e., $q_A\stackrel{0}{\oplus}q_B=q_A+q_B$. The $\kappa$-sum  has the following properties: (1) it is associative, where
$(q_A\stackrel{\kappa}{\oplus}q_B)\stackrel{\kappa}{\oplus}q_C
=q_A\stackrel{\kappa}{\oplus}(q_B\stackrel{\kappa}{\oplus}q_C)$;
(2) it admits a neutral element, where $q_A\stackrel{\kappa}{\oplus}0=0
\stackrel{\kappa}{\oplus}a_A=q_A$; (3) it admits an opposite element, where
$q_A\stackrel{\kappa}{\oplus}(-q_A)=(-q_A) \stackrel{\kappa}{\oplus}q_A=0$;
(4) it is commutative, where
$q_A\stackrel{\kappa}{\oplus}q_B=q_B\stackrel{\kappa}{\oplus}q_A$. 
Then, the algebraic structure $({\bf
R},\stackrel{\kappa}{\oplus})$ forms an abelian group. The $\kappa$-difference
$\stackrel{\kappa}{\ominus}$ is defined as
$q_A\stackrel{\kappa}{\ominus}q_B=q_A\stackrel{\kappa}{\oplus}(-q_A)$. 

Starting from the $\kappa$-sum, $\kappa$-mathematics can be introduced after defining the $\kappa$-exponential function as the solution to the functional Equation (\ref{20Y-III-24}). The introduction of $\kappa$-functions can be performed starting from the $\kappa$-exponential and following the standard procedures of ordinary mathematics. For instance, $\kappa$-trigonometry (ordinary or hyperbolic) can be introduced by employing the $\kappa$-Euler formula, while the $\kappa$-inverse function follows after the inversion of their direct functions~\cite{PhA2001}. Also, $\kappa$-differential calculus can be introduced after defining the $\kappa$-derivative as the differential operator, which acts on the $\kappa$-exponential function, which subsequently produces the $\kappa$-exponential function itself. 

Next, we revisit the $\kappa$-derivative and discuss how it arises within the special relativity. Let us consider two identical particles $A$ and $B$ in the one-dimension spatial frame $\cal {S}$ having dimensionless momenta $q_A=q$ and $q_B={\Tilde{q}}$, respectively. In the rest frame ${\cal S}'$ of particle $B$, which is an inertial frame that moves with velocity $v_B$ with respect the inertial frame ${\cal S}$,  the dimensionless moment of particle $B$ is $q'_B={\Tilde {q}}'=0$, while the dimensionless moment $q'_A=q'$ of particle $A$ is given by $q'= q \stackrel{\kappa}{\ominus} \Tilde{q}$. We suppose that $\Tilde{q} \approx q$ and pose $dq \approx q-\Tilde{q}$ and $dq'\approx q \stackrel{\kappa}{\ominus} \Tilde{q}$. Starting from the limit

\begin{eqnarray}
\lim_{\Tilde{q}  \rightarrow q}\,\frac{q \stackrel{\kappa}{\ominus} \Tilde{q} }{q-\Tilde{q} }= \frac{1}{\gamma_{\kappa}(q)} \  \label{20Y-III-33}
\end{eqnarray}
with $\gamma_{\kappa}(q)=\sqrt{1+\kappa^2 q^2}$ being the Lorentz factor, the differential $d q'$ can be obtained as 
\begin{eqnarray}
d q' =\frac{dq}{\gamma_{\kappa}(q)} \  \label{20Y-III-34}
\end{eqnarray}
{The} $\kappa$-differential $d _{\kappa}q= d q'$ has a very transparent physical meaning representing the infinitesimal variation in the momentum of a given particle, observed in the frame ${\cal S}'$. It is related to the infinitesimal variation in the momentum $dq$ of the same particle, observed in the inertial frame ${\cal S}$ through the Lorentz factor. A further interesting property of the differentials $d _{\kappa}q$ is given by $d _{\kappa}q= d (\rho_{\kappa}(q)) $ or simply $d _{\kappa}q= d \rho$, where $\rho=\rho_{\kappa}(q)$ is the $\kappa$-rapidity defined through  
\begin{eqnarray}
\rho_{\kappa} (q) = \frac{1}{\kappa} {\rm arcsinh} ( \kappa \,q ) \ 
\label{20Y-III-35} 
\end{eqnarray}
{The} variable $\phi_{\kappa}(u)= {\rm arctanh} (v/c) ={\rm arctanh} (\kappa u)$ was introduced into special relativity in 1910 by V. Varicak and E. T. Whittak and was named rapidity by A. Robb in 1911. The old rapidity is related to the $\kappa$-rapidity $\rho_{\kappa} (q)$ through  $\phi_{\kappa}(u) =\kappa\,\rho_{\kappa} (q)$, which can be easily verified after taking into account that $u=q/\gamma_{\kappa}(q)$. The presence of the proportionality factor $\kappa$ in the relation linking $\phi_{\kappa}(u)$ and $\rho_{\kappa} (q)$ is not trivial because, in the classical limit, the $\kappa$-rapidity reduces to the dimensionless momentum, i.e., $\rho_{0} (q)=q$, while in the same limit, the old rapidity does not reduce to the dimensionless velocity holding $\phi_{0}(u)=0$. The relativistic composition law of $\kappa$-rapidity is given by
\begin{eqnarray}
\rho_{\kappa} (q'_A) = \rho_{\kappa} (q_A) - \rho_{\kappa} (q_B) \ 
\label{20Y-III-36} 
\end{eqnarray}
and becomes identical to the ordinary difference $\rho'_A = \rho_A - \rho_B$. The expression of the $\kappa$-exponential function in terms of $\rho_{\kappa} (q)$ is given by 
\begin{eqnarray}
\exp_{\kappa}(q)= \exp (\rho_{\kappa} (q))  \ 
\label{20Y-III-37} 
\end{eqnarray}

The $\kappa$-derivative of the scalar function $f(q)$ is defined through
\begin{equation}
\frac{d\,f(q)}{d_{\kappa} \,q}= \gamma_{\kappa}(q) \, \frac{d\,f(q)}{d \,q} \ \  \label{20Y-III-38}
\end{equation}
{It} is important to note that $d\,f(q)$ is an ordinary differential, while $d_{\kappa} \,q$ is a $\kappa$-differential. It follows that the $\kappa$-derivative is proportional through the Lorentz factor $\gamma_{\kappa}(q)$ to the ordinary derivative and then obeys Leibniz's rules of the ordinary derivative.

\subsection{The $\kappa$-Differential Equations}

The dynamic variables of relativistic physics can be obtained as solutions of first-order differential equations involving the $\kappa$-derivative $d f(q)/d_{\kappa}q$, which, in the classical limit, reduces to the corresponding differential equations of classical physics.

The solution to
\begin{eqnarray}
\frac{d}{d_{\kappa}q}\, f(q)=1  \  \label{20Y-III-39}
\end{eqnarray}
with the condition $f(0)=0$ is the rapidity function $f(q)=\rho_{\kappa}(q)$, i.e., $f(q)=\frac{1}{\kappa} {\rm arcsinh} (\kappa q)$.

The solution to
\begin{eqnarray}
\frac{d}{d_{\kappa}q}\, f(q)= q  \  \label{20Y-III-40}
\end{eqnarray}
with the condition that $f(0)=1/\kappa^2$ is the total energy  $f(q)={\cal E}_{\kappa}(q)$, i.e., $f(q)=\sqrt{1+\kappa^2 q^2}/\kappa^2$, while the solution to the same equation with the condition $f(0)=0$ is the relativistic kinetic energy $f(q)={\cal W}_{\kappa}(q)$ i.e., $f(q)=\left (\sqrt{1+\kappa^2 q^2} -1 \right )/\kappa^2$. 

The solution to
\begin{eqnarray}
\frac{d}{d_{\kappa}q}\, f(q)=\kappa^2 q  \  \label{20Y-III-41}
\end{eqnarray}
with the condition $f(0)=1$ is the Lorentz factor $f(q)=\gamma_{\kappa}(q)$, i.e., $f(q)=\sqrt{1+\kappa^2 q^2}$.

The solution to
\begin{eqnarray}
\frac{d}{d_{\kappa}q}\, f(q)= f(q)  \  \label{20Y-III-42}
\end{eqnarray}
with the condition $f(0)=1$ is the $\kappa$-exponential function $f(q)= \left ( \sqrt{1+\kappa^2 q^2} + \kappa q \right)^{1/\kappa}$.

Finally, the relativistic velocity $u_{\kappa}(q)=q/\sqrt{1+\kappa^2q^2}$ is the solution $f(q)=u_{\kappa}(q)$ of the differential equation
\begin{eqnarray}
\frac{d}{d_{\kappa}q}\, f(q)= \left (\frac{f(q)}{q}\right )^2  \  \label{20Y-III-43}
\end{eqnarray}
with the condition $f(\pm \infty)=\pm 1/\kappa$.

\subsection{The Scaling Property of ${\kappa}$-Logarithm}

The differential equation 
\begin{eqnarray}
\sqrt{1+\kappa^2 q^2} \,\, \frac{d \,\exp_{\kappa}\,(q)}{dq}=\exp_{\kappa}(q) \
\  \label{20Y-III-44}
\end{eqnarray}
obeyed by the $\exp_{\kappa}(q)$ can be easily inverted, obtaining 
\begin{eqnarray}
\frac{d \,\ln_{\kappa}\,(w)}{dw}=\frac{\gamma_{\kappa}(\ln_{\kappa}(w))}{w}  \
\  \label{20Y-III-45}
\end{eqnarray}
and after taking into account Equation~(\ref{20Y-III-27}), it follows that the $\kappa$-logarithm function obeys the first-order differential--functional equation
\begin{eqnarray}
\frac{d}{dw}\,[\,w\,\,\ln_{\kappa}\,(w)\,]=\frac{1}{\gamma} \,\ln_{\kappa}\left(\epsilon w\right) \
\  \label{20Y-III-46}
\end{eqnarray}
expressing the so-called scaling property of the $\kappa$-logarithm. In the  $\kappa \rightarrow 0$ classical limit, the latter equation continues to hold and reduces to a well-known property of the ordinary logarithm, where scaling constants reduce to the values $\gamma=1$ and $\epsilon=e$. 

The two last equations, if combined, lead to the further property of 
$\kappa$-logarithm
\begin{equation}
\frac{d^2}{dw^2}\,[\,w\,\ln_{\kappa}\,(w)]=\frac{1}{\gamma \, w} \,\gamma_{\kappa}(\ln_{\kappa}\left(\epsilon w\right) \geq 0 \ \  \label{20Y-III-47}
\end{equation}

\section{\boldmath{$\kappa$}-Statistical Physics}\label{sec4}

\subsection{Maximum Entropy Principle and $\kappa$-Entropy}

In proposing a relativistic statistical theory, the only guiding principle available is the metaphor of classical statistical physics, and the entropy form plays an important role in this context. The standard relativistic statistical theory is based on an entropic form identical to that of classical statistical physics, the BGS entropy. This is due to the great success of BGS entropy in classical many-body physics. In Einstein's special relativity, all microscopic physical quantities such as particle momentum or particle energy are properly generalized. Regarding macroscopic quantities such as temperature or pressure, there is still a debate about how they should be defined in a relativistic context. It is, therefore, an evident dichotomy that on one side, there is the BGS entropy, which dominates both classical and relativistic physics, and on the other side, there are all the other physical quantities, both microscopic and macroscopic, which are or could be modified in special relativity. We note that the BGS entropy in the relativistic context conducts to the Juttner distribution, which, when considered as a function of the relativistic particle energy, is exactly the Boltzmann exponential factor of classical physics. It has long been known that the Boltzmann factor does not correctly describe the spectrum of cosmic rays, which are relativistic particles.

In the following, we will present the relativistic statistical theory based on the $\kappa$-entropy $S_{\kappa}$, which is defined as the standard mean of the opposite of the $\kappa$-logarithm emerging in special relativity. The paradigm of classical statistical physics will be constantly present in our discussion, and the starting point will be the maximum entropy principle, the cornerstone of statistical theory. Let us consider the constrained entropy $\Phi(f)=S_{\kappa}(f)+{\cal C}(f)$, where the constraints functional ${\cal C}(f)$ is given in its simplest form by
\begin{eqnarray}
{\cal C}(f)={\rm a_{1}} \left[\,\sum_{i}\, f_i-1 \right] + {
a_2}\left[{  I } - \sum_{i}\, \, { I_i}\,f_i \right] \, 
\label{20Y-IV-48}
\end{eqnarray}
where $a_1$ and $a_2$ are the Lagrange multipliers, while $\{I_i\}$ is the generator function of the moment $I=\sum_i I_i f_i$. The variational equation $\frac{\delta \Phi(f)}{\delta f_i}=0$ implies the maximization of $S_{\kappa}$ under the constraints imposing the conservation of the norm of $f_i$ and the a priori knowledge of the values of the moment $I=\sum_i I_i f_i$ generated by the generator function $\{I_i\}$. The solution to the above variational problem conducts to the equation $\frac{d}{d f_i}(f_i \ln_{\kappa} (f_i))=a_1-a_2 I_i$, which, after taking into account the scaling axiom, assumes the form 
\begin{eqnarray}
\frac{1}{\gamma}\, \ln_{\kappa} (\epsilon f_i) = a_1 - a_2 I_i \, \,  
\label{20Y-IV-49}
\end{eqnarray}
{In} the case of a classical particle gas, the above equation reduces to $\ln (ef_i)=a_1 -a_2 I_i$, where the microscopic collisional invariant $I_i$ is the classical particle energy in the state $i$, while Lagrange multipliers are related to the gas temperature $T$ and chemical potential $\mu$ according to $a_2=1/k_B T$ and $a_1=\mu/k_B T$. The same parameters $T$ and $\mu$ will also occur in the case of a relativistic gas, while $I_i$ will be the microscopic relativistic collisional invariant. Equation~(\ref{20Y-IV-49}), after inversion, takes the form 
\begin{equation}
f_i= \frac{1}{\epsilon} \exp_{\kappa} \left (\!-\frac{I_i - \mu}{k_B T_{\kappa}}\, \right )  \ \  \label{20Y-IV-50}
\end{equation}
with $T_{\kappa}= T/ \gamma$.

Remarkably, thanks to the scaling property of the $\kappa$-logarithm, the expressions of the $\kappa$-entropy and the $\kappa$-distribution $f_i$ are given in terms of the same function, which appears in its direct ($\kappa$-logarithm) or inverse ($\kappa$-exponential) form, just as in the classical case.

Let us pose $w_i=(I_i-\mu)/k_B T_{\kappa}$. When $w_i \rightarrow + \infty$, the asymptotic behavior of the function $\exp_{\kappa}(-w_i)$ is given by $\exp_{\kappa}(-w_i) \approx (2\kappa w_i)^{- 1/\kappa}$. Consequently, the tail of the distribution (\ref{20Y-IV-50}) is described by a Pareto power law function, i.e., $f_i\approx \epsilon^{-1}(2\kappa w_i)^{-1/\kappa}$, instead of the exponential tails of the Juttner distribution $f_i=e^{-1} \exp(-w_i)$, originating from the BGS entropy. The power-law tail of the distribution (\ref{20Y-IV-50}) is one of its most interesting features and is consistent with the experimental evidence in relativistic particle physics, i.e., cosmic rays and the so-called $\kappa$-plasmas observed in laboratory or in astrophysics.

Let us introduce the $\kappa$-entropy $S_{\kappa}(g)=\sum_i \sigma(g_i)$, which refers to the arbitrary distribution $g=\{g_i\}$ and is subjected to the constraints described by the functional
${\cal C}(g)= -\sum_{i}\, ( {\rm a_{2}}I_i - {\rm a_{1}})\, g_i -{\rm a_{1}} + {\rm a_{2}} I$. Let us further denote by $f=\{f_i\}$ the optimal distribution defined according to the maximum entropy principle, defined in Equation~(\ref{20Y-IV-49}), which takes the form $a_2 \, I_i - {a}_{1}=\frac{d \, \sigma(f_i) }{d \,f_i}$, so that the constraints functional can be written as follows:
\begin{eqnarray}
{\cal C}(g)= -{\rm a_{1}} + {\rm a_{2}} I  - \sum_{i}\, \frac{d \, \sigma(f_i) }{d \, f_i} \, g_i   \, 
\label{20Y-IV-51}
\end{eqnarray}

The difference in the constrained entropy $\Phi(g)$ from its maximum value $\Phi(f)$, i.e., $\Phi(f)-\Phi(g)=S_{\kappa}(f)-S_{\kappa}(g) + {\cal C}(f) -{\cal C}(g)$, finally assumes the form 
\begin{equation}
{\Phi}(f)\!-\!{\Phi}(g)=\sum_i\left[\,
\sigma(f_i)\!-\!\sigma(g_i)\!-\!\frac{d \,
\sigma(f_i)}{d \, f_i}(f_i\!-\!g_i)\right]  \  \label{20Y-IV-52}
\end{equation}
{When} $g_i\approx f_i$, the Taylor expansion can be considered as
\begin{equation}
\sigma(g_i)\approx \sigma(f_i)+ \frac{d \,
\sigma(f_i)}{d \, f_i}(g_i-f_i) + \frac{1}{2}
\frac{d^2 \sigma(f_i)}{d \, f_i^{\,2}}(g_i-f_i)^{2} 
\label{20Y-IV-53}
\end{equation}
so that we obtain
\begin{eqnarray}
{ \Phi}(f)-{\Phi}(g)\approx -\sum_i \frac{1}{2} \frac{ d ^2
\sigma(g_i)}{d\, g_i^{\,2}}(f_i-g_i)^2  \  \label{20Y-IV-54}
\end{eqnarray}
and after taking into account the expression of $\frac{ d ^2
\sigma(g_i)}{d\, g_i^{\,2}}$ as given by Equation~(\ref{20Y-III-47}), we obtain
\begin{eqnarray}
{ \Phi}(f)-{\Phi}(g)\approx \sum_i  \frac{\gamma_{\kappa}(\ln_{\kappa}(\epsilon g_i))}{2\,\gamma \, g_i}(f_i-g_i)^2  \geq 0 \ \label{20Y-IV-55}
\end{eqnarray}  
{The} latter relationship tells us that $\Phi(f)$ represents the maximum value of $\Phi(g)$ and expresses the thermodynamic stability of the system.

Another stability that differs from thermodynamic stability is the Lesche stability condition, which prescribes that any physically meaningful entropy that depends on a probability distribution function $g$ should exhibit a small relative error
\begin{equation}
R=\left|\frac{S(g)-S(h)}{S_{max}}\right| \  \label{20Y-IV-56}
\end{equation}
with respect to small changes in the probability distributions $g \rightarrow h$
\begin{equation}
D=||g-h|| \  \label{20Y-IV-57}
\end{equation}
{Mathematically}, this means that for every $\varepsilon>0$, there is a $\delta>0$ so that $R\leq \varepsilon$ applies to all distribution functions that fulfill $D\leq \delta$. It is known that the Lesche stability condition holds for the Boltzmann--Shannon entropy, and in refs.~\cite{KSPA04,AKSJPA04}, it was shown that the Lesche stability condition also holds for the $\kappa$-entropy. In addition, the $\kappa$-entropy is also Lesche-stable in the thermodynamic limit.

\subsection{$\kappa$-Kinetics}

Let us consider the first equation of the Bogoliubov--Born--Green--Kirkwood--Yvon hierarchy, which describes the evolution of a relativistic many-body system in the presence of an external force field and imposes particle conservation during collisions~\cite{PRE2005,Groot,Cercignani}:
\begin{eqnarray}
p^{\,\nu}\partial_{\nu}f-m F^{\nu}\frac{\partial f}{\partial p^{\,\nu}}= \!\int \!\!\!\!\!\!\! &&
\frac{d^3p'}{{p'}^{0}}\frac{d^3p_1}{p_1^{\,0}}\frac{d^3p'_1}{{p'}_{\!\!1}^{0}}
\,\,G \nonumber \\ &&\times \left[C(f', f'_1)-C(f, f_1)\right]
\label{20Y-IV-58} \ \ \ \ \ \ \ \ \ \ 
\end{eqnarray}
{The} system is described by the one-particle correlation function or distribution function $f=f(x,p)$, where $x$ and $p$ are the four-vector position and momentum. In the above equation, both the streaming term and the Lorentz invariant integrations in the collision integral have the standard forms of relativistic kinetic theory. The two-particle correlation function $C(f,f_1)$, which is determined below, is postulated in the case of ordinary relativistic kinetics as $C(f,f_1)=f\,f_1$, which represents the molecular chaos hypothesis and reduces the above evolution equation to the relativistic Boltzmann equation.

Following standard lines of kinetic theory, we note that in
stationary conditions, the collision integral vanishes and then $C(f, f_1)= C(f', f'_1)$. This relationship expresses a conservation law for the particle system and must have the form $L(f)+ L(f_{1}) =L(f^\prime) +L(f^\prime_{1})$. In relativistic kinetics, the collision invariant
$L(f)$, unless an additive constant, is proportional to the microscopic relativistic invariant $I(x,p)$, i.e.,
\begin{equation}
L(f)=- a_2 I(x,p) + a_1 \ 
\label{20Y-IV-59}
\end{equation}
with $a_1$ and $a_2$ being two arbitrary constants. The more general
microscopic relativistic invariant $I$, in the presence of an external electromagnetic field $A^{\nu}$, has a form proportional~to
\begin{equation}
I(x,p)=\left(p^{\nu}+q A^{\nu}\!/c \right)\,U_{\nu}-mc^2 \ 
\label{20Y-IV-60}
\end{equation}
with $U_{\nu}$ being the hydrodynamic four-vector velocity with
$U^{\nu}U_{\nu}=c^2$~\cite{Groot}. 

The expression of the distribution function defined in Equation~(\ref{20Y-IV-50}) holds in stationary conditions where the entropy of the particle system reaches its maximum value. According to the scaling axiom, after considering the correspondences $f_i \rightarrow f(x,p)$ and $I_i \rightarrow I(x,p)$ and after the identification of $a_2=1/k_B T$ and $a_1=\mu/k_B T$, it follows that $ L(f) = \lambda_{\kappa} (f) $, or more explicitly
\begin{equation}
L(f)= \frac{1}{\gamma} \, \ln_{\kappa} (\epsilon f)   \ \ 
\label{20Y-IV-61}
\end{equation}
while the stationary distribution assumes the form
\begin{equation}
f(x,p)=\frac{1}{\epsilon} \exp_{\kappa}\left (-\frac{I(x,p) -\mu}{k_BT_{\kappa}} \,\right ) \  \label{20Y-IV-62}
\end{equation}
with $T_{\kappa}=T / \gamma$.

\subsection{$\kappa$-Molecular Chaos Hypothesis}

In stationary conditions, $C(f, f_1)=C(f', f'_1)$ applies. This relationship expresses a conservation law and can be written in the form $L(f)+ L(f_{1}) =L(f^\prime) +L(f^\prime_{1})$ after posing  
\begin{equation}
C(f, f_1)=  L^{-1} ( \, L (f) + L (f_1) )   \ \ 
\label{20Y-IV-63}
\end{equation}
The function $L(w)$ increases monotonically on the interval $0\leq w < + \infty$, with $L(0)=-\infty$ and $L(+\infty)=+\infty$. These conditions imply that $C(0,f_1)=C(f,0)=0$, just as in the case of the ordinary correlation function. After taking into account the expression of the function $L(f)$, the two-particle correlation function assumes the form
\begin{equation}
\epsilon \, C(f, f_1)=  \exp_{\kappa} ( \,\ln_{\kappa} (\epsilon f) + \ln_{\kappa} (\epsilon f_1) )   \ \ 
\label{20Y-IV-64}
\end{equation}
which can be written in a more compact form 
\begin{equation}
\epsilon \, C(f, f_1)=  (\epsilon f)\otimes (\epsilon f_1)   \ \ 
\label{20Y-IV-65}
\end{equation}
by involving the generalized product 
\begin{eqnarray}
g\otimes h= \exp_{\kappa} ( \,\ln_{\kappa} g + \ln_{\kappa} h) \ \  \label{20Y-IV-66}
\end{eqnarray}
This $\kappa$-product between probabilities has the following properties:

\noindent (i) $(g\otimes
h)\otimes l=g\otimes (h\otimes l)$, i.e., it is associative;

\noindent (ii) $g\otimes h=h \otimes g$, i.e., it is commutative;

\noindent (iii)  $1\otimes g=g$, i.e., it admits the unity as a neutral element; 

\noindent (iv) $g\otimes(1/g)=1$, i.e., the inverse element of $g$ is $1/g$;

\noindent (v)  It holds the property $g\otimes 0=0$;

\noindent (vi)  $g{\oslash}h=g\otimes(1/h)$ defines the $\kappa$-division between probabilities.

\noindent The real, positive probability distribution functions form an abelian group. The properties of $\otimes$ are the same as those of the ordinary product, so the two products are isomorphic.

We can conclude that the relation given by Equation~(\ref{20Y-IV-64}), which defines the two-particle correlation function by the $\kappa$-product, is the relativistic version of the molecular chaos hypothesis and reduces to its standard form $C(f, f_1) \propto f\,f_1$ in the classical limit $\kappa \rightarrow 0$.

\subsection{Four-Vector $\kappa$-Entropy and Relativistic H-Theorem}

In standard relativistic kinetics, it is known from the H-theorem that entropy production is never negative and that there is no entropy production under equilibrium conditions. In the following, we will demonstrate the H-theorem for the system governed by the kinetic Equation (\ref{20Y-IV-58}). We define the four-vector entropy $S^{\nu}=(S^{0},\mbox{\boldmath $S$})$ as follows:
\begin{equation}
S^{\nu}= -\int \frac{d^3p}{p^{0}}\,p^{\nu}\, \, f \, \ln_{\kappa} (f)
\  \label{20Y-IV-67}
\end{equation}
and note that the scalar entropy $S^{0}=S_{\kappa}$ coincides with the $\kappa$-entropy, while $\mbox{\boldmath $S$}=\mbox{\boldmath $S$}_{\kappa}$ is the $\kappa$-entropy flow. After considering the identity
$d^3p/p^0=d^4p \,\, 2\,
\theta(p^0)\,\delta(p^{\mu}p_{\mu}-m^2c^2)$ and the observation that
$d^4p$ is a scalar because the Jacobian of the
Lorentz transformation is equal to unity, we conclude that $S^{\nu}$
transforms as a four-vector, since $p^{\nu}$
transforms as a four-vector.

In order to calculate the entropy production $\partial_{\nu}S^{\nu}$, we start from the definition of $S^{\nu}$ and the relationship $\partial_{\nu} \, [f \, \ln_{\kappa} (f)]=[\,\partial \,[ f \, \ln_{\kappa} (f)]/\,\partial f \,] \, \partial_{\nu}f= \lambda_{\kappa} (f) \,\partial_{\nu} f$ with $\lambda_{\kappa}(f)=\frac{1}{\gamma} \ln_{\kappa}(\epsilon f)$,~obtaining
\begin{eqnarray}
\partial_{\nu}S^{\nu}= -  \int
\frac{d^3p}{p^{0}}\, \lambda_{\kappa}(f)\,\,p^{\nu}\,\partial_{\nu}f  \label{20Y-IV-68} \ \ 
\end{eqnarray}
After taking into account the kinetic equation (\ref{20Y-IV-58}), the entropy production assumes the~form
\begin{eqnarray}
\partial_{\nu}S^{\nu} \!=\! -\!\int
\frac{d^3p}{p^{0}}\frac{d^3p'}{{p'}^{0}}\frac{d^3p_1}{p_1^{\,0}}\frac{d^3p'_1}{{p'}_{\!\!1}^{0}}
\,\,G \, \lambda_{\kappa} (f) \nonumber \\ \times \left[ C(f', f'_1)\!-\! C(f, f_1) \right]   \nonumber \\ -  m \int
\frac{d^3p}{p^{0}}\, \lambda_{\kappa} (f)\,\,F^{\nu}\frac{\partial
f}{\partial p^{\,\nu}}     \label{20Y-IV-69}
\end{eqnarray}

Since the Lorentz force $F^{\nu}$ has the properties
$p^{\nu}F_{\nu}=0$ and $\partial F^{\nu}/\partial p^{\nu}=0$, the
last term in the above equation involving $F^{\nu}$ is equal to
zero~\cite{Groot}.
Given the particular symmetry of the non-vanishing integral in Equation~(\ref{20Y-IV-69})
we can write the entropy production as follows

\begin{eqnarray}
\partial_{\nu}S^{\nu} = \frac{1}{4} \int  \frac{d^3p}{p^{0}}
\frac{d^3p'}{{p'}^{0}}\frac{d^3p_1}{p_1^{\,0}}\frac{d^3p'_1}{{p'}_{\!\!1}^{0}} 
\,G \left[C(f', f'_1)\!- \!C(f, f_1)\right] \nonumber \\ \times  [ \lambda_{\kappa}(f') \!+ \!\lambda_{\kappa}(f'_1)\! -\! \lambda_{\kappa}(f)\! -\! \lambda_{\kappa}(f_1) ] \ \ \  \ \ \ \ \ \ \  \label{20Y-IV-70}
\end{eqnarray}

From the definition of the two-particle correlation function, it follows that $\lambda_{\kappa}(f') + \lambda_{\kappa}(f'_1) - \lambda_{\kappa}(f) - \lambda_{\kappa}(f_1)= \lambda_{\kappa}(C(f', f'_1)) - \lambda_{\kappa}(C(f, f_1)) $, and after posing $\alpha'=C(f', f'_1)$, $\alpha=C(f, f_1)$, finally, we write Equation~() 
in the form
\begin{eqnarray}
\partial_{\nu}S^{\nu}=\frac{1}{4} \int \!\!\!\!\!\!\!&&\frac{d^3p}{p^{0}}
\frac{d^3p'}{{p'}^{0}}\frac{d^3p_1}{p_1^{\,0}}\frac{d^3p'_1}{{p'}_{\!\!1}^{0}}
\,\,G \nonumber \\ &&\times \left [\alpha'- \alpha \right] [\lambda_{\kappa}(\alpha') - \lambda_{\kappa}(\alpha) ] \ \  \ \ \ \ \ \ \ \label{20Y-IV-71}
\end{eqnarray}
with  $\lambda_{\kappa}(\alpha)$ being an increasing function, it follows that $\left [\alpha'- \alpha \right] [\lambda_{\kappa}(\alpha') - \lambda_{\kappa}(\alpha) ] \geq 0$, $\forall \alpha', \alpha$, and then we can conclude that
\begin{equation}
\partial_{\nu}S^{\nu}\geq 0 \ \  \label{20Y-IV-72}
\end{equation}
This last relation is the local formulation of the relativistic
H-theorem, which represents the second law of thermodynamics
for the system governed by the evolution Equation (\ref{20Y-IV-58}).

\subsection{Relativistic Temperature}

The construction of a thermodynamic theory compatible with the principles of special relativity is an old and still open problem, dating back to the first years immediately after the proposal of the relativistic theory. The proposals that have dealt with the question of how the thermodynamic quantities that characterize the physical system change when the inertial reference system changes are diverse and contradictory. Some of these proposals are still under consideration, and the problem is still highly topical. Let $T$ denote the temperature of a body at rest and $T'$ the temperature of the body when the body is observed from a reference frame moving at a speed characterized by the Lorentz factor $\gamma$. According to Planck and Einstein, the two temperatures are linked by $T'=T/\gamma$. According to Ott, $T'= \gamma T$. Finally, according to Landsberg, $T'=T$. In a series of subsequent articles, some of which have appeared recently, the scientific community has overwhelmingly sided with the Planck--Einstein proposal and accepted that a moving body is colder.

We do not intend to go into this important topic here. However, it is noteworthy that the present formalism proves to be consistent with the Planck--Einstein proposal. Let us consider the particle gas described by the distribution function (\ref{20Y-IV-62}), where the relevant temperature is given by
\begin{equation}
T_{\kappa}= \frac{1}{\gamma} \, T \ \  \label{20Y-IV-73}
\end{equation}
with
\begin{equation}
\gamma=\frac{1}{\sqrt{1-\kappa^2}}=\frac{1}{\sqrt{1-(\frac{v^*}{c})^2}} \ \  \label{20Y-IV-73}
\end{equation}
A possible interpretation of the above formula of temperature is the following. The temperature of the system at rest is $T_0=T$ while $T_{\kappa} = T_0/\gamma < T_0$ is its temperature when it is moving at speed $v^*$; this is just the Planck--Einstein proposal. The two temperatures $T_{\kappa}$ and $T_0$ are proportional to each other, and the proportionality factor is the Lorentz factor $\gamma$. Remarkably, entropy does not have this proportionality property. The entropy of the physical system moving at the speed $v^*$ is $S_{\kappa}=-\sum_i \, f_i\, \ln_{\kappa}(f_i)$, while its entropy at rest reduces to the classical Boltzmann entropy $\,S_{0}=\,-\sum_i\, f_i \, \ln \,(f_i)$.

\section{Epilogue}\label{sec5}

Some of the results of the present theory, which were discussed in the previous sections, are emphasized below:

\noindent
{ {(i) Relativistic statistical theory:}} It is possible to construct a statistical theory within the framework of special relativity that preserves the main features of classical statistical theory (axiomatic structure, maximum entropy principle, thermodynamic stability, Lesche stability, molecular chaos hypothesis, local formulation of H-theorem, etc.).

\noindent
{ {(ii) Old problems of special relativity:}} Within the framework of the new relativistic statistical theory, answers naturally arise to questions that were formulated immediately after the proposal of special relativity as to how the temperature and entropy of a moving body change. In particular, it turns out that the temperature varies according to the law $T_{\kappa}=T_0 /\gamma$ proposed by Planck and Einstein in 1906, where $\gamma=1/\sqrt{1-\kappa^2}$ is the Lorentz~factor.

\noindent
{{(iii) Axiomatic structure of the theory:}} Although the statistical theory generated by the entropy $S_{\kappa}$ was developed within the framework of Einstein's special relativity, it can also be introduced without reference to special relativity given its applications outside physics by following the guidelines of information theory, which emphasizes the axiomatic structure of the various theories. In the construction of $\kappa$-entropy, the first three Khinchin--Shannon axioms are taken into account, i.e., those of the continuity, maximality, and expansibility of the ordinary Boltzmann entropy. Subsequently, the fourth Khinchin--Shannon axiom of strong additivity is replaced by two new axioms, namely, those of self-duality and scaling, which express well-known properties of logarithmic Boltzmann entropy. In the final step, it is shown that these five axioms are not only able to generate the Boltzmann entropy but also a further and unique entropy, namely, $\kappa$-entropy, which turns out to be a one-parameter continuous generalization of the Boltzmann entropy. The axioms of self-duality and scaling can be seen as stemming from the first principles of special relativity. In any case, these two axioms can also be easily justified outside the special relativity, since they have general validity and can also generate the Boltzmann entropy.

\noindent
{{(iv) $\kappa$-mathematical statistics:}} Statistical theory does not only include statistical mechanics, which is a physical theory. Mathematical statistics is another important tool for analyzing complex systems. Two important families of distributions dominate ordinary mathematical statistics. On the one hand, there is the family of distributions with exponential tails (generalized gamma distribution, Weibull distribution, logistic distribution, etc.), and on the other hand, the family of distributions with power-law tails (Pareto, Log-Logistic, Burr type XII or Singh-Maddala distribution, Dagum distribution, etc.). This dichotomy can be overcome in the framework of the present formalism by using the $\kappa$-exponential function instead of the ordinary exponential function in the construction of statistical distributions, obtaining a unique family of statistical distributions ($\kappa$-generalized gamma distribution, $\kappa$-Weibull distribution, $\kappa$-logistic distribution, etc.). The new unified class of $\kappa$-distributions~\cite{EPL2021} in the low spectral region reproduces the standard family of exponential distributions, while in the high spectral region, it exhibits Pareto power-law tails.

\noindent
{ {(v) $\kappa$-mathematics:}} In special relativity, the physical quantities such as momentum, kinetic energy, etc. are relativistically generalized and change their expressions relatively to the corresponding classical expressions. The composition laws of the various physical quantities are also properly generalized. The generalized sum of relativistic moments inevitably leads to the generalization of the entire mathematics. The resulting $\kappa$-calculus allows for the introduction of relativistic functions such as the $\kappa$-exponential, the $\kappa$-logarithm, the $\kappa$-trigonometry, and so on. $\kappa$-mathematics proves to be isomorphic to ordinary mathematics, which classically obtains the $\kappa \rightarrow 0$ limit.

\noindent
{{(vi) The Gell-Mann plectic:}} $\kappa$-mathematics is based on a formalism that can handle both simple systems (relativistic one-particle physics) and complex systems (relativistic statistical physics). Furthermore, the same formalism makes it possible to treat physical and non-physical complex systems (statistical physics, information theory, and statistical mathematics) in a unified way. The above features of the $\kappa$-formalism give it the status of a candidate for the construction of the holistic theory of simple and complex systems, called {\it {plectics}
} by Gell-Mann~\cite{GellMann1,GellMann2}.

\vspace{6pt}

\end{document}